\documentclass[12pt,a4paper,notitlepage]{article}

\usepackage{amssymb,amsmath}

\usepackage{indentfirst}

\usepackage[large,bf]{caption2}

\newcommand{\bsl}{\boldsymbol}

\begin{document}

\large

\author{I.M.Narodetskii$^{1)}$, C.Semay$^{2)}$,\\
B.Silvestre-Brac$^{3)}$, Yu.A.Simonov$^{1)}$, M.A.Trusov$^{1)}$}

\title{Pentaquarks in the Jaffe--Wilczek approximation}

\date{}

\maketitle

\begin{abstract}
\noindent The masses of $uudd\bar s $, $uudd\bar d$ and $uuss\bar
d$ pentaquarks are evaluated in a framework of  both the Effective
Hamiltonian approach to QCD and spinless Salpeter using the
Jaffe--Wilczek diquark approximation and the string interaction
for the diquark--diquark--antiquark system. The pentaquark masses
are found  to be in the region above 2 GeV. That indicates that
the Goldstone boson exchange effects may play an important role in
the light pentaquarks. The same calculations yield the mass of
$[ud]^2\bar c$ pentaquark $\sim$ 3250 MeV and $[ud]^2\bar b$
pentaquark $\sim$ 6509 MeV.
\end{abstract}

\vspace*{70mm}

\footnoterule

\vspace*{10mm}

{\small\noindent $^{1)}$ Institute of Theoretical and Experimental
Physics, Moscow, Russia.\\ $^{2)}$ Groupe de Physique Nucl\'eaire
Th\'eorique, Universit\'e de Mons-Hainaut, Acad\'{e}mie
universitaire Wallonie-Bruxelles, Mons, Belgium.\\ $^{3)}$
Laboratoire de Physique Subatomique et de Cosmologie,
Grenoble-Cedex, France.}

\newpage

\section{Introduction}

Recently LEPS and DIANA collaborations
\cite{Nakano:2003,Barmin:2003} reported the observation of a very
narrow peak in the $K^+n$ and $K^0p$ invariant mass distribution
which existence has been confirmed by several experimental groups
in various reaction channels \cite{2prim}. These experimental
results were motivated by the pioneering paper on chiral soliton
model \cite{Diakonov:1997}. The reported mass determinations for
the $\Theta$ are very consistent, falling in the range
1540$\pm$10, with the width smaller than the experimental
resolution of $20$ MeV for the photon and neutrino induced
reactions and of 9 MeV for the ITEP $K^+ \mathrm{Xe} \to K^0 p
\mathrm{Xe}'$ experiment.

From the soliton point of view the $\Theta$ is nothing exotic
compared with other baryons -- it is just a member of
${\bf\overline{10}}_F$  multiplet with $S=+1$. However, in the
sense of the quark model $\Theta^+(1540)$ baryon with positive
amount of strangeness is manifestly exotic -- its minimal
configuration can not be satisfied by three quarks. The positive
strangeness requires an $\bar s$ and $qqqq$ (where $q$ refers to
the lightest quarks $(u,d)$) are required for the net baryon
number, thus making a pentaquark $uudd\bar s$ state as the minimal
``valence'' configuration. Later  NA49 collaboration  at CERN SPS
\cite{NA49} announced evidence for an additional  narrow $udus\bar
s$ resonance with $I=3/2$, a mass
 $1.862\pm 0.002$ GeV and a width below the detector resolution
of about 18 MeV\footnote{NA49 also reports evidence for a
$\Xi^0$(1860) decaying into $\Xi(1320)\pi$.} and H1 collaboration
at HERA \cite{H1} found a narrow resonance
 in $D^{*-}p$ and $D^{*+} \bar p$ invariant mass combinations
at $3.099 \pm 0.033_{\text{stat}} \pm 0.005_{\text{syst}}$ GeV and
a measured Gaussian width of $12 \pm 3_{\text{stat}}$ MeV,
compatible with the experimental resolution. The later resonance
is interpreted as an anti-charmed baryon with a minimal
constituent quark composition of $uudd\bar c$, together with the
charge conjugate. The discoveries of first manifestly exotic
hadrons mark the beginning of a new and rich spectroscopy in QCD
and provide an opportunity to refine our quantitative
understanding of nonperturbative QCD at low energy.

The $\Theta$-hyperon has hypercharge $Y=2$ and third component of
isospin $I_3=0$. The apparent absence of the $I_{3}=+1$,
$\Theta^{++}$ in $K^{+}p$ argues against $I=1$, therefore it is
usually assumed the $\Theta$ to be an isosinglet. The other
quantum numbers are not established yet.

As to the theoretical predictions we are faced with a somewhat
ambiguous situation, in which exotic baryons may have been
discovered, but there are important controversies with theoretical
predictions for masses of pentaquark states. The experimental
results triggered a vigorous theoretical activity and put a
renewed urge in the need to understand of how baryon  properties
are obtained from QCD.

All attempts of the theoretical estimations of the pentaquark
masses can be subdivided into following four categories: (i)
dynamical calculations using the sum rules or lattice QCD
\cite{sumrules,Fodor},
 (ii) the
phenomenological analyses of the hyperfine splitting in quark
model \cite{stancu,Karliner}, (iii) phenomenological analyses of
the $SU(3)_F$ mass relations, and (iv) dynamical calculations
using the chiral $SU(3)$ quark model \cite{chiral_quark_model}).

The QCD sum rules predict a negative parity $\Theta^+$ of mass
$\simeq 1.5$ GeV, while no positive parity state was
found~\cite{sumrules}. The lattice QCD study also predicts that
the parity of the lowest $\Theta$ hyperon is most likely negative
\cite{Fodor}.

The naive quark models, in which all constituents are in a
relative $S$-wave, naturally predict the ground state energy of a
$J^P=\frac{1}{2}^-$ pentaquark to be lower than that of a
$J^P=\frac{1}{2}^+$ one. However, using the arguments based on
both the Goldstone boson exchange between constituent quarks and
color-magnetic exchange it was mentioned that the increase of
hyperfine energy in going from negative to positive parity states
can be quite enough to compensate the orbital excitation energy
$\sim 200$ MeV. However, existing dynamical calculations of
pentamasses using the chiral $SU(3)$ quark model (see, e.g.,
\cite{chiral_quark_model}) are subject to significant
uncertainties and can not be considered as conclusive.

Pentaquark baryons are unexpectedly light. Indeed, a naive quark
model with quark mass $\sim$ 350 MeV predicts $\Theta^+$ at about
\(350\times5=1750\) MeV plus $\sim$ 150 MeV for strangeness plus
$\sim$ 200 MeV for the $P$-wave excitation. A natural remedy would
be to decrease the number of constituents. This leads one to
consider dynamical clustering into subsystems of diquarks like
$[ud]^2\bar s$ \cite{Jaffe:2003} and/or triquarks like
$[ud][ud\bar s]$ \cite{Karliner} which amplify the attractive
color-magnetic forces. In particular, in \cite{Jaffe:2003} it has
been proposed that the systematics of exotic baryons can be
explained by diquark correlations.

The quark constituent model have not been yet derived  from QCD.
Therefore it is tempting to consider the Effective Hamiltonian
(EH) approach in QCD (see, e.g.,  \cite{lisbon}) which from one
side can be derived from QCD and from another side leads to the
results for the $\bar q q$ mesons and $3q$ baryons which are
equivalent to the quark model ones with some important
modifications. The EH approach contains the minimal number of
input parameters: current (or pole) quark masses, the string
tension $\sigma$ and the strong coupling constant $\alpha_s$, and
does not contain fitting parameters as, e.g., the total
subtraction constant in the Hamiltonian. It should be useful and
attractive to consider expanding of this approach to include
diquark degrees of freedom with appropriate interactions. The
preview of this program was done in \cite{NTS03}. It is based on
assumption that chiral and the short range gluon exchange forces
are responsible for the formation of $ud$ diquarks in $\Theta$
while the strings are mainly responsible for binding constituents
in $\Theta$. In this paper we review and extend application of the
EH approach to the Jaffe--Wilczek model of pentaquarks.

In this model inside $\Theta(1540)$ and other $q^{4}\bar q$
baryons the four quarks are bound into two scalar, singlet isospin
diquarks. Diquarks must couple to ${\bf 3}_c$ to join in a color
singlet hadron. In the quark model  five quarks are connected by
seven strings. In the diquark approximation the short legs on this
figure  shrink to points and the five-quark system effectively
reduces to the three-body one, studied within the EH approach in
\cite{baryons,trusov}. In total there are six flavor symmetric
diquark pairs $[ud]^2$, $[ud][ds]_+$, $[ds]^2$, $[ds][su]_+$,
$[su]^2$, and $[su][ud]_+$ combining with the remaining antiquark
which give 18 pentaquark states in ${\bf 8}_F$ plus
${\bf\overline{10}}_F$.  All these states are degenerate in the
SU(3)$_F$ limit.

\section{The EH approach and the results}

The EH for the three constituents has the form
\begin{equation}
\label{EH} H=\sum\limits_{i=1}^3\left(\frac{m_i^{2}}{2\mu_i}+
\frac{\mu_i}{2}\right)+H_0+V,
\end{equation}
where $H_0$ is the kinetic energy operator, $V$ is the sum of the
perturbative one-gluon exchange potentials and the string
potential $V_{\rm{string}}$.

The dynamical masses $\mu_i$ (analogues of the constituent ones)
are expressed in terms of the current quark masses $m_i$ from the
condition of the minimum of the hadron mass $M_H^{(0)}$ as
function of $\mu_i$ \footnote{~Technically, this is done using the
auxiliary field approach to get rid of the square root term in the
Lagrangian \cite{polyakov,brin77}. Applied to the QCD Lagrangian,
this technique yields the EH  for hadrons (mesons, baryons,
pentaquarks) depending on auxiliary fields $\mu_i$. In practice,
these fields are finally treated as $c$-numbers determined from
(\ref{minimum_condition}).}:
\begin{equation} \label{minimum_condition}
\frac{\partial M_H^{(0)}(m_i,\mu_i)}{\partial \mu_i}=0, ~~~
M_H^{(0)}=\sum\limits_{i=1}^3\left(\frac{m_i^{2}}{2\mu_i}+
\frac{\mu_i}{2}\right)+E_0(\mu_i), \end{equation} $E_0(\mu_i)$
being eigenvalue of the operator $H_0+V$. Quarks acquire
constituent masses $\mu_i\sim\sqrt{\sigma}$ due to the string
interaction in (\ref{EH}). As of today the EH in the form of
(\ref{EH}) does not include chiral symmetry breaking effects. A
possible interplay with these effects should be carefully
clarified in the future.

The physical mass $M_H$ of a hadron is
\begin{equation}\label{self_energy}
M_H=M_H^{(0)}+\sum_i C_i.
\end{equation}
The (negative) constants $C_i$ have the meaning of the constituent
self energies and are explicitly expressed in terms of string
tension $\sigma$ \cite{simonov_self_energy}:
\begin{equation}\label{c_i}
C_i=-\frac{2\sigma}{\pi\mu_i}\eta_i,\end{equation} where
\begin{equation} \label{eta_i} \eta_q=1,~~ \eta_s=0.88, ~~\eta_c=0.234,~~\eta_b=0.052. \end{equation} In Eq.
(\ref{eta_i}) $\eta_s$, $\eta_c$, and $\eta_b$  are the correction
factors due to nonvanishing current masses of the strange, charm
and bottom quarks, respectively. The self-energy corrections are
due to constituent spin interaction with the vacuum background
fields and equal zero for any scalar constituent.

Accuracy  of the EH method for the three-quark systems is $~\sim
100$ MeV or better \cite{baryons,trusov}. One can expect the same
accuracy for the diquark--diquark--(anti)quark system.

Consider a pentaquark consisting of two identical diquarks with
current mass $m_{[ud]}$ and antiquark with current mass $m_{\bar
q}$ ($q=d,s,c$). In the hyperspherical formalism the wave function
$\psi( {\bsl\rho},{\bsl\lambda)}$ expressed in terms of the Jacobi
coordinates $\bsl \rho$ and $\bsl \lambda$ and can be written in a
symbolical shorthand as
\begin{equation}\psi(\bsl{\rho},\bsl{\lambda})=\sum\limits_K\psi_K(R)Y_{[K]}(\Omega),
\end{equation}where $Y_{[K]}$ are eigen functions (the
hyperspherical harmonics) of the angular momentum operator $\hat
K(\Omega)$ on the 6-dimensional sphere:
$\hat{K}^2(\Omega)Y_{[K]}=-K(K+4)Y_{[K]}$, with $K$ being the
grand orbital momentum. For identical diquarks, like $[ud]^{2}$,
the lightest state must have a wave function antisymmetric under
diquark space exchange. There are two possible pentaquark wave
functions antisymmetric under diquark exchange, the first one
(with lower energy) corresponding to the total orbital momentum
$L=1$, and the second one (with higher energy) corresponding to
$L=0$. For a state with $L=1,~l_{\rho}=1,~l_{\lambda}=0$ the wave
function in the lowest hyperspherical approximation $K=1$ reads:
\begin{equation}
 \psi=R^{-5/2}\chi_1(R)u_1(\Omega),~~~
 u_1(\Omega)=\sqrt{\frac{8}{\pi^2}}\sin\theta\cdot
 Y_{1m}(\Hat{\bsl{\rho}}),
\end{equation} where $R^2=\bsl{\rho}^2+\bsl{\lambda}^2$. Here one
unit of orbital momentum between the diquarks is with respect to
the $\bf\rho$ variable whereas the ${\bf\lambda}$ variable is in
an $S$-state.  The Schr\"odinger equation for $\chi_1(R)$ written
in terms of the variable $x=\sqrt{\mu} R$, where $\mu$ is an
arbitrary scale of mass dimension which drops off in the final
expressions, reads:
\begin{equation} \label{shr}
\frac{d^2\chi_1(x)}{dx^2}+
2\left[E_0+\frac{a_1}{x}-b_1x-\frac{35}{8x^2}\right] \chi_1(x)=0,
\end{equation}
with the boundary condition $\chi_K(x) \sim {\cal O} (x^{7/2})$ as
$x\to 0$ and the asymptotic behavior $\chi_1(x)\sim
{\mathrm{Ai}}((2b_1)^{1/3}x)$ as $x\to \infty$. In Eq. (\ref{shr})
\begin{equation}
\begin{aligned}
a_1&=R\sqrt{\mu}\cdot \int
V_{\text{C}}(\bsl{r}_1,\bsl{r}_2,\bsl{r}_3)\cdot u_1^2\cdot
d\Omega,\\ b_1&=\frac{1}{R\sqrt{\mu}}\cdot\int
V_{\text{string}}(\bsl{r}_1,\bsl{r}_2,\bsl{r}_3)\cdot u_1^2\cdot
d\Omega,
\end{aligned} \label{ab_int}
\end{equation}
where
\begin{equation}
V_{\text{C}}(\bsl{r}_1,\bsl{r}_2,\bsl{r}_3)=
-\frac{2}{3}\alpha_s\cdot\sum\limits_{i<j}\frac{1}{r_{ij}},
\end{equation}
and
\begin{equation}
V_{\text{string}}(\bsl{r}_1,\bsl{r}_2,\bsl{r}_3)=\sigma\cdot
l_{\text{min}}
\end{equation}
is proportional to the total length of the strings, i.e., to the
sum of the  distances of (anti)quark or diquarks from the string
junction point. In the Y-shape, the strings meet at $120^\circ$ in
order to insure the minimum energy. This shape moves continuously
to a two-legs configuration where the legs meet at an angle larger
than $120^\circ$. Explicit expression of
$V_{\text{string}}(\bsl{r}_1,\bsl{r}_2,\bsl{r}_3)$ in terms of
Jacobi variables is given in \cite{plekhanov}.

The mass of the $\Theta^+$ obviously depends on $m_{[ud]}$ and
$m_s$. The current masses of the light quarks are relatively
well-known: $m_{u,d}\approx 0$, $m_s\approx 170$ MeV. The only
other parameter of strong interactions is the effective mass of
the diquark $m_{[ud]}$. In principle, this mass could be computed
dynamically. Instead, one can tune $m_{[ud]}$ (as well as
$m_{[us]}$ and $m_{[ds]}$) to obtain the baryon masses in the
quark--diquark approximation. We shall comment on this point later
on.

In what follows, we use $\sigma=0.15\text{~GeV}^2$, and explicitly
include the Coulomb-like interaction between quark and diquarks
with $\alpha_s=0.39$.

For the pedagogy, let us first assume  $m_{[ud]}=0$. This
assumption leads to the lowest $uudd\bar d$ and $uudd\bar s$
pentaquarks. If the current diquark masses vanish, then the
$[ud]^2\bar d$ pentaquark is dynamically exactly analogous to the
$J^P=\frac{1}{2}^-$ nucleon resonance and $[ud]^2\bar s$
pentaquark is an analogue of  the $J^P=\frac{1}{2}^-$ $\Lambda$
hyperon, with one important exception. The masses of $P$-wave
baryons calculated using the EH method acquire the (negative)
contribution $3C_q$ for $J^P=\frac{1}{2}^-$ nucleons  and
$2C_q+C_s$ for the $J^P=\frac{1}{2}^-$ hyperons. These
contributions are  due to the interaction of constituent spins
with the vacuum chromomagnetic field. Using the results of Table 1
below  we get the mass of the $P$-wave nucleon resonance with the
orbital $\rho$ excitation 1600 MeV and the mass of $\Lambda$
hyperon 1600 MeV that within 100 MeV agrees with the known
$P$-wave $N$ and $\Lambda$ resonances.

However, the above discussion shows that the self-energies
$C_{[ud]}$ equal zero for the scalar diquarks. This means that
introducing any scalar constituent increases the pentaquark energy
(relative to the $N$ and $\Lambda$ $P$-wave resonances) by
$2|C_q|\sim 200-300$ MeV. Therefore prior any calculation we can
put the lower bound for the pentaquark in the Jaffe--Wilczek
approximation, $M(\Theta)\ge 2\text{~GeV}$.

The numerical calculation for $m_{[ud]}=0$ yields the mass of
$[ud]^2\bar s$ pentaquark $\sim$ 2100 MeV (see Table
\ref{table1}). The similar calculations yield the mass of
$[ud]^2\bar c$ pentaquark $\sim$ 3250 MeV (for $m_c=1.4$ GeV) and
$[ud]^2\bar b$ pentaquark $\sim$ 6509 MeV (for $m_b=4.8$
GeV)\cite{veselov}. For illustration of accuracy of the auxiliary
field (AF) formalism in Table \ref{table1} are also shown the
masses of $[ud]^2\bar d$ and $[ud]^2\bar d$ pentaquarks calculated
using the spinless Salpeter equation (SSE):
\begin{eqnarray*} H_S
&=& \sum_{i=1}^3 \sqrt{\bsl{p}_i^2+ m_i^2} + V, \\ M &=& M_0 -
\frac{2\sigma}{\pi} \sum_{i=1}^3 \frac{\eta_i}{\left<
\sqrt{\bsl{p}_i^2+ m_i^2}  \right>},
\end{eqnarray*}
where $V$ is the same as in Eq. (\ref{EH}), $M_0$ is the
eigenvalue of $H_S$,
$\left<\sqrt{\bsl{p}_{[ud]}^2+m_{[ud]}^2}\right>$,
$\left<\sqrt{\bsl{p}_{\bar q}^2+m_{\bar q}^2}\right>$ are the
average kinetic energies of diquarks and an antiquark and $\eta_i$
are the correction factors given in (\ref{eta_i}). The numerical
algorithm to solve the three-body SSE is based on an expansion of
the wave function in terms of harmonic oscillator functions with
different sizes \cite{nunb77}. In fact to apply this techniques to
the three-body SSE we need to use an approximation of the
three-body potential $V_{{\rm string}}$ by a sum of the two- and
one-body potentials, see \cite{NSSB}. This approximation, however,
introduces the marginal correction to the energy eigenvalues. The
quantities $\mu_{[ud]}$ and $\mu_q$ denote either the constituent
masses calculated in the AF formalism using Eq. (\ref{c_i}) or
$\left<\sqrt{\bsl{p}_{[ud]}^2+m_{[ud]}^2}\right>$,
$\left<\sqrt{\bsl{p}_{\bar q}^2+m_{\bar q}^2}\right>$ found from
the solution of SSE. It is seen from Table 1 that these quantities
agree with accuracy better than 5$\%$. The pentaquark masses
calculated by the two methods differ by 100 MeV  for
$([ud]^2\bar{s})$ and 160 MeV for $[ud]^2\bar{d}$. The
approximation of $V_{{\rm string}}$ mentioned above introduces the
correction to the energy eigenvalues $\le 30$ MeV, so we conclude
that the results obtained using the AF formalism and the SSE agree
within $\sim 5\%$, i.e., the accuracy of the AF results for
pentaquarks is the same as for the $q{\bar q}$ system (see, e.g.,
\cite{MNS}).

If we withdraw an assumption $m_{[ud]}=0$, then a possible  way to
estimate the current diquark masses is to tune $m_{[ud]}$,
$m_{[us]}$ and $m_{[ds]}$ from the fit to the nucleon and hyperon
masses (in the quark--diquark approximation). In this way one
naturally obtains larger pentaquark masses.  We have performed
such the calculations using the SSE. We briefly investigated the
sensitivity of the pentaquark mass predictions to the choice of
$\sigma$, the strange quark mass $m_s$ and diquark masses
$m_{[ud]}$ and found $M([ud]^2\bar d)$ in the range 2.2--2.4 GeV,
$M([ud]^2\bar s)$ $\sim$  2.4 GeV and $M([us]^2\bar d)$ $\sim$ 2.5
GeV.

Increasing $\alpha_s$ up to $0.6$ (the value used in the
Capstick--Isgur model \cite{capstick-isgur}) decreases the
$[ud]^2\bar s$ mass by $\sim$ $120$ MeV (see Table \ref{table2}).
We have briefly investigated the effect of the hyperfine
interaction due to the $\sigma$ meson exchange between diquarks
and strange antiquark and found that it  lowers the $\Theta^+$
energy by $\sim 180$ MeV for $g^2_{\sigma}/4\pi\sim 1$. As the
result we obtain the lower bound of $[ud]^2\bar s$ pentaquark
$M([ud]^2\bar s)=1740$ MeV  (for $m_{[ud]}=0$) which is still
$\sim$ 200 MeV above the experimental value.

\section{Conclusions}

We therefore conclude that the string dynamics alone in its
simplified form predicts too high masses of pentaquarks. This may
indicate on a large role of the chiral symmetry breaking effects
in light pentaquark systems. An ``extremal'' approach of chiral
soliton model totally neglects the confinement effects and
concentrates on the pure chiral properties of baryons. Therefore
the existence of $\Theta$, if confirmed, provides an unique
possibility to clarify the interplay between the quark and chiral
degrees of freedom in light baryons.


This work was supported by RFBR grants No. 03-02-17345,
04-02-17263, the grant for leading scientific schools No.
1774.2003.2. The  NATO is also greatly acknowledged for the grant
No. PST.CLG.978710.

\newpage

\section*{Tables}

\renewcommand{\captionlabeldelim}{.}

\begin{table}[h]
\caption{The pentaquark masses in the quark--diquark--diquark
approximations.  The masses of $[ud]^2\bar{q}$ states ($q=d,s$)
for $J^P=\frac{1}{2}^+$ pentaquarks are shown in units of GeV.}
\begin{center}
\large
\begin{tabular}{|c|c|c|c|c|}
\hline & & \(\mu_{[ud]}\) & \(\mu_q\) & \(M\) \\ \hline & AF &
0.482 & 0.458 & 2.171 \\ \(\smash{[ud]^2\bar{s}~\dfrac{1}{2}^+}\)
& & & &
\\ & SSE & 0.463 & 0.468 & 2.070
\\ \hline & AF & 0.476 & 0.415 & 2.091 \\ \(\smash{[ud]^2\bar{d}~\dfrac{1}{2}^+}\) & & & &
\\ & SSE & 0.469 & 0.379 & 1.934
\\ \hline
\end{tabular}
\end{center}
\label{table1}
\end{table}


\begin{table}[h]
\caption{The $[ud]^2\bar s$ mass calculated using the SSE as a
function of $\alpha_s$ with and without the Goldstone boson
exchange (GBE) in units of GeV.}
\begin{center}
\large
\begin{tabular}{|c|c|c|c|}
\hline \(\alpha_s\) & 0.39 & 0.50 & 0.60 \\ \hline \(M\) {\small
(with GBE)} & 1.893 & 1.814 & 1.737 \\ \hline \(M\) {\small
(without GBE)} & 2.069 & 2.007 & 1.949 \\ \hline
\end{tabular}
\end{center}
\label{table2}
\end{table}

\end{document}